\documentclass[12pt]{article}
\usepackage{amssymb}
\usepackage{graphicx}

\begin{document}

\title{Electroproduction of hadrons on nuclei at GeV energies}
\author{T.~Falter\footnote{e-mail: Thomas.Falter@theo.physik.uni-giessen.de} 
~and U.~Mosel\\ 
Institut fuer Theoretische Physik\\ 
Universitaet Giessen\\ 
D-35392 Giessen, Germany}
\date{\today}
\maketitle

\begin{abstract}
We investigate coherence length effects and hadron attenuation 
in lepton scattering off nuclei in the kinematic regime of the HERMES 
experiment. The elementary electron-nucleon interaction is described within 
the event generator PYTHIA while a full coupled-channel treatment of the 
final state interactions is included by means of a BUU transport model. The 
results of our calculations are in good agreement with the experimentally 
measured transparency ratio of incoherent $\rho^0$ electroproduction off 
$^{14}$N and the multiplicity ratio $R_M^h$ for charged hadrons, pions, kaons, protons and anti-protons in deep inelastic scattering off $^{14}$N and $^{84}$Kr targets. 
\end{abstract}

\bigskip

\section{Introduction}

High energy meson electroproduction off complex nuclei offers a promising 
tool to study the physics of hadron formation. The relatively clean 
nuclear environment of electron induced reactions makes it possible
to investigate the timescale of the hadronization process as well as the 
properties of hadrons immediately after their creation. In addition one can 
vary the energy and virtuality of the exchanged photon to examine the 
phenomenon of color transparency \cite{CT}. 

In previous works \cite{Eff00,Fal02,Fal03a} we have developed a 
method to combine the quantum mechanical coherence in the entrance channel 
of photonuclear reactions with a full coupled channel treatment of the final 
state interactions (FSI) in the framework of a semi-classical transport 
model. This allows us to include a much broader class of final state 
interactions than usual Glauber theory. 

In \cite{Eff00,Fal02} we have shown that
side feeding effects that are absent in calculation based on Glauber theory 
might play an important role in semi-inclusive photoproduction 
experiments. In \cite{Fal03a} we have then investigated the transparency 
ratio of exclusive incoherent $\rho^0$ electroproduction off $^{14}$N and $^{84}$Kr 
within our model and found excellent agreement with the Nitrogen 
data~\cite{HERMESrho} from the HERMES collaboration. The most important 
elementary process for this 
reaction was found to be diffractive $\rho^0$ production on bound nucleons. 
Other production mechanisms were supressed by the stringent exclusivity 
measure imposed by experiment. We found no signature for color transparency in the data for the transparency ratio 
\begin{equation}
  T_A=\frac{\sigma_{\gamma^*A\rightarrow \rho^0A^*}}{A\sigma_{\gamma^*N\rightarrow \rho^0N}}
\end{equation}
as a function of the $\rho^0$ 
coherence length, meaning that in our model calculations we could assume 
that the diffractively produced $\rho^0$ starts to interact with a hadronic 
cross section right after its production. However, Kopeliovich et al.
\cite{Kop01} have pointed out that the missing signal of color transparency 
might be 
an accidental consequence of the specific correlation between the $Q^2$ of 
the photon and the coherence length of the $\rho^0$ in the HERMES data. 

A finite formation time of hadrons becomes visible in our investigation \cite{Fal03b} of the HERMES data on 
charged particle multiplicities in deep inelastic lepton scattering (DIS) off 
nuclei \cite{HERMESDIS_N,HERMESDIS_Kr}. In the DIS regime we assume that the primary 
production mechanism is governed by the excitation and decay of a hadronic 
string. The fragmentation of the string and the hadronization of the 
fragments takes some time (formation time $\tau_f$). After their formation 
time the reaction products undergo hadronic final state interactions with 
nucleons in the target nucleus. Due to time dilatation the formation time $t_f$ in the target rest frame can become quite large:
\begin{equation}
\label{eq:formation-time}
        t_f=\gamma\cdot\tau_f=\frac{z_h\nu}{m_h}\cdot\tau_f .
\end{equation}
Here $m_h$ denotes the hadron mass and $z_h$ is the fraction of the
photon energy $\nu$ that is carried away by the hadron. For high energies and for very light fragments, e.g. pions, the formation length can easily exceed the size of a nucleus. This, however, is not the case for the heavier fragments such as baryons and vector mesons for which the 
formation length at HERMES energies is still of the order of the nuclear 
radius so that ``classical'' nuclear interactions must play a role for these 
particles. In our model the prehadronic interactions of the constituent quarks from the 
string ends during the formation time are effectively accounted for 
by using the concept of leading hadrons which is in 
accordance with other transport models for high energy reactions 
such as the UrQMD \cite{URQMD2} and the HSD model \cite{Geiss,Cas99}. 
Alternative explanations for the observed modification of the 
multiplicity spectra in terms of a possible rescaling of the fragmentation 
function in nuclei \cite{Acc02} or a purely partonic energy loss \cite{Wang,Arl03} through multiple scattering of the struck quark and induced gluon bremsstrahlung may model these prehadronic interactions, but may neglect the 
classical FSI after formation.

The formation time also plays an important 
role in studies of ultrarelativistic heavy ion reactions. For example, the 
observed quenching of high transverse momentum hadrons in $Au+Au$ reactions 
relative to $p+p$ collisions is often thought to be due to jet quenching in a 
quark gluon plasma. However, the attenuation of high $p_T$ hadrons might also 
be due to hadronic rescattering processes \cite{Gal02} if the hadron 
formation time $\tau_f$ (in its rest frame) is sufficiently short. 

Our paper is structured in the following way: In Sec. \ref{subsec:shadowing} 
we show how we describe the electron-nucleon interaction and how we account 
for shadowing within our model. The transport 
model is sketched in Sec. \ref{subsec:transport}. Our result for the 
transparency ratio of incoherent $\rho^0$ electroproduction is presented 
in Sec. \ref{subsec:coherence} in comparison with experimental data from the 
HERMES experiment. In Sec. \ref{subsec:formation} we show the calculated
multiplicity ratio for charged hadrons, pions, kaons, protons and anti-protons. We close with a short summary in Sec. \ref{sec:summary}.


\section{Model}
\label{sec:model}
In our approach the lepton-nucleus interaction is splitted
into two parts: 1) In the first step the virtual photon is absorbed
on a bound nucleon of the target; this interaction produces a
bunch of particles that in step 2) are propagated within the
transport model. Coherence length effects in the entrance channel,
that give rise to nuclear shadowing, are taken into account as
described in Ref.~\cite{Fal02}.

\subsection{Elementary photon nucleon interaction and shadowing}
\label{subsec:shadowing}
We use the event generator PYTHIA~v6.2~\cite{PYTHIA} to describe the 
interaction of the virtual photon and a nucleon. The basic idea is that in 
a photon hadron collision the physical photon $\gamma$ not necessarily 
interacts as a point particle $\gamma_0$ (direct interaction) but might
fluctuate into a vector meson 
$V=\rho^0,\omega,\phi,J/\Psi$ (vector meson dominance, VMD) or perturbatively branch
into a $q\bar{q}$ pair before the interaction (generalized vector meson 
dominance, GVMD). We have shown in \cite{Fal03a} that the latter is very 
unlikely in the kinematic regime of the HERMES experiment (photon energy 
$\nu\approx 7-23.4$~GeV, $Q^2\approx 0.5-5$~GeV$^2$). In most cases the photon 
will therefore interact directly with a parton of the target nucleon 
(DIS) or fluctuate into a vector meson (VMD) before it reaches 
the nucleon. In the latter case the vector meson might either scatter 
diffractively from the nucleon or a hard scattering between the constituents of
the vector meson and the nucleon 
might take place. The hard scattering like the direct photon nucleon 
interaction leads to the excitation of one or more hadronic strings which 
finally fragment into hadrons. 

If the struck nucleon is embedded in a nucleus one has to account for its Fermi
motion and binding energy as well as for Pauli blocking of final state nucleons.
In addition one has to be aware that the nuclear environment influences
the VMD part of the photon nucleon interaction, since the vector meson 
components get modified on their way through the nuclear medium to the 
interaction point. The latter has been investigated in detail in 
Ref.~\cite{Fal03a}. There we have shown that at position $\vec r$ inside the 
nucleus the physical photon state has changed to 
\begin{equation}
\label{eq:inmediumstate}
  |\gamma(\vec r)\rangle=\left(1-\sum_{V=\rho,\omega,\phi,J/\Psi}\frac{e^2}{2g_V^2}F_V^2\right)|\gamma_0\rangle+\sum_{V=\rho,\omega,\phi,J/\Psi}\frac{e}{g_V}F_V\left(1-\overline{\Gamma^{(A)}_V}(\vec r)\right)|V\rangle
\end{equation}
due to these 'initial state interactions' of the photon. The formfactor $F_V$ from Ref. \cite{Fri00} accounts also for contributions of longitudinal 
photons. If one neglects any influence of the FSI for the moment the 
reaction amplitude for the process $\gamma N\rightarrow f$ on a nucleon at 
position $\vec r$ inside a nucleus changes compared to the vacuum in the
following way:
\begin{equation}
  \label{eq:transition}
  \langle f|\hat{T}|\gamma\rangle\rightarrow\langle f|\hat{T}|\gamma(\vec r)\rangle .
\end{equation}
The nuclear profile function $\Gamma^{(A)}_V(\vec r)$ can be expressed as
\begin{eqnarray}
  \overline{\Gamma^{(A)}_V}(\vec b,z)&=&\intop_{-\infty}^{z}dz_in(\vec b,z_i)(1-j_0(q_c|z_i-z|))\frac{\sigma_{VN}}{2}(1-i\alpha_V)e^{iq_V(z_i-z)} \nonumber\\
& &\times\exp\left[-\frac{1}{2}\sigma_{VN}(1-i\alpha_V)\intop_{z_i}^{z}dz_kn(\vec b,z_k)\right]  \label{eq:av_profile}
\end{eqnarray}
with the nucleon number density $n(\vec r)$, the total vector meson nucleon
cross section $\sigma_{VN}$ and the ratio $\alpha_V$ of real to imaginary part 
of the vector meson nucleon forward scattering amplitude. The Bessel function parametrization with $q_c=0.78$~GeV avoids unphysical 
contributions from processes where $z_i\approx z$ which would 
contribute for small values of the coherence length $l_V$. In 
Ref.~\cite{Fal00} we have shown that this Bessel function parameterization yields 
a good description of shadowing in photoabsorption down to the onset region
of shadowing.

The probability that the photon interacts with a bound nucleon via a certain 
vector meson fluctuation therefore depends on the position inside the nucleus.
The momentum transfer 
\begin{equation}
  q_V=\sqrt{\nu^2+Q^2}-\sqrt{\nu^2-m_V^2}
\end{equation}
in the phase factor of Eq. \ref{eq:av_profile} equals the inverse of the 
coherence length $l_V$ of a vector meson component $V$, i.e. the length
that the photon travels as a vector meson fluctuation estimated via the
uncertainty principle.


\subsection{Transport model}\label{subsec:transport}
The propagation of the produced final state $|f\rangle$ through the nucleus is treated 
within a semi-classical transport model based on the 
Boltzmann-Uehling-Uhlenbeck (BUU) equation.
The BUU equation describes the time evolution of the phase space density 
$f_i(\vec r,\vec p,t)$ of particles of type $i$ that can interact via binary 
reactions. Besides the nucleons these particles involve baryonic 
resonances and mesons ($\pi$, $\eta$, $\rho$, $K$, ...) that are produced 
either in the primary reaction or during the FSI. For a particle species $i$
the BUU equation can be written as:
\begin{equation}
  \left(\frac{\partial}{\partial t}+\frac{\partial H}{\partial\vec r}\frac{\partial}{\partial \vec r}-\frac{\partial H}{\partial \vec r}\frac{\partial}{\partial \vec p}\right)f_i(\vec r,\vec p,t)=I_{coll}[f_1,...f_i,...,f_M].
\end{equation}
For baryons the Hamilton function $H$ includes a mean field potential which 
depends on the particle position and momentum. The collision integral on the 
right hand side accounts for the creation and
annihilation of particles of type $i$ in a collision as well as elastic 
scattering from one position in phase space into another. For fermions Pauli 
blocking is taken into account in $I_{coll}$ via blocking factors. The BUU 
equations of each particle species $i$ are coupled via the mean 
field and the collision integral. The resulting system of coupled 
differential-integral equations is solved via a test particle ansatz for 
the phase space density. We also stress that within our model instable
particles might decay during their propagation through the nucleus. For 
details of the transport model see Ref.~\cite{Eff99}. 

The classes of FSI that are included in the transport model go
far beyond what can be achieved within Glauber theory. As a result the finally
observed particles do not need to be created in the primary reaction but might
be produced during the FSI via side feeding. It is therefore clear that a 
purely absorptive treatment of the FSI as in Glauber theory can only be used if
one is sure that one has eliminated the possibility of side feeding by 
applying enough constraints on the observable (see~\cite{Fal02} for details).
This might be possible for exclusive $\rho^0$ production measured at HERMES
but becomes questionable for the measurement of inclusive hadron multiplicity 
spectra. 

\begin{figure}[bt]
  \begin{center}
       \includegraphics[width=8cm]{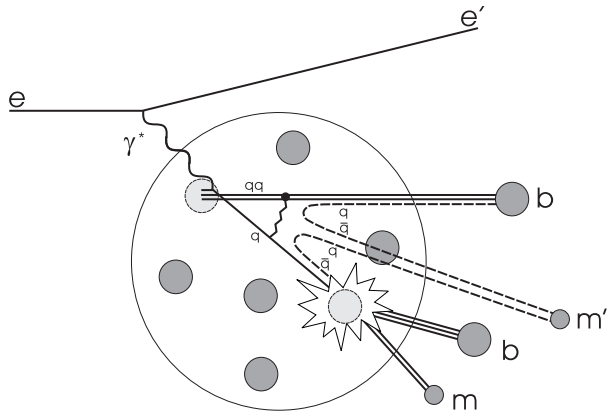}
  \end{center}
  \vspace{-0.5cm}
  \caption{Illustration of an electron nucleus interaction: The virtual 
photon $\gamma^*$ excites a hadronic string by hitting a quark $q$ inside a 
bound nucleon. In our example the string between the struck quark $q$ and 
diquark $qq$ fragments due to the creation of two quark-antiquark pairs.
One of the antiquarks combines with the struck quark to form a 'leading' 
meson $m$, one of the created quarks combines with the diquark to form a 
'leading' baryon $b$. The remaining partons combine to a meson $m'$ that, 
depending on the mass of the meson, might leave the nucleus before
it hadronizes (see (\ref{eq:formation-time})).}
  \label{fig:fig1}
\end{figure}

As discussed in Sec. \ref{subsec:shadowing} a hard primary photon nucleon 
interaction leads to the excitation of hadronic strings. The time, that is 
needed for the fragmentation of the strings and for the hadronization of the
fragments, we denote as formation time $\tau_f$ in line with the
convention in transport models. For simplicity we assume that the
formation time is a constant $\tau_f$ in the rest frame of each
hadron and that it does not depend on the particle species. In
principle, one expects a distribution in the formation times,
however, we here address only the mean value of such an 'unknown'
distribution. We recall, that due to time dilatation the formation
time $t_f$ in the laboratory frame is then proportional to the
particle's energy as can be seen from Eq. (\ref{eq:formation-time}).
The size of $\tau_f$ can be estimated by the time that the constituents of the
hadrons need to travel a distance of a typical hadronic radius
(0.5--0.8~fm). We assume that hadrons, whose constituent quarks
and antiquarks are created from the vacuum in the string
fragmentation, do not interact with the surrounding nuclear medium
within their formation time. For the leading hadrons, i.e. those
involving quarks (antiquarks) from the struck nucleon or the hadronic
components of the photon, we assume a
reduced effective cross section $\sigma_{lead}$ during the
formation time $\tau_f$ and the full hadronic cross section $\sigma_h$ \cite{PDG} later
on. This concept is illustrated in Fig.~\ref{fig:fig1}: due to
time dilatation light particles emerging from the middle of
the string might escape the nucleus without further interaction if
they carry a large fraction $z_h$ of the photon energy $\nu$ (7~GeV 
$\leq \nu \leq 23.4 $~GeV). However, the hadrons with $z_h$
close to 1 (or energy $\approx \nu$) predominantly stem from the
string ends and therefore can interact directly after the
photon-nucleon interaction. We note, that about $2/3$ of the
intermediate $z_h$ hadrons (mainly pions) are created from the
decay of vector mesons that have been created in the string
fragmentation. Because of their higher mass $m_h$ (0.77 -- 1.02~GeV) 
these vector mesons may form (or hadronize) inside the
nucleus (see Eq. (\ref{eq:formation-time})) and thus suffer from
FSI. The effect of the final-state interactions, finally, will
depend dominantly on the nuclear geometry, i.e. the size of the
target nucleus. 


\section{Results}

\subsection{Incoherent $\rho^0$ production}
\label{subsec:coherence}
Incoherent $\rho^0$ electroproduction off nuclei has been studied in
Ref.~\cite{Fal03a}. In the following we will summarize the results
of our investigations.
For our model calculations we demand that the final state has to consist of two oppositely charged pions plus an ensemble of bound nucleons and we 
use the kinematic cuts of the HERMES collaboration~\cite{HERMES99}. 
This means that we apply the exclusivity measure
\begin{equation}
  \label{eq:exclusivity}
  -2\textrm{~GeV}<\Delta E=\frac{p_Y^2-m_N^2}{2m_N}<0.6\textrm{~GeV},
\end{equation}
where $m_N$ denotes the nucleon mass and 
\begin{equation}
  \label{eq:py}
  p_Y=p_N+p_{\gamma}-p_{\rho}
\end{equation}
the 4-momentum of the undetected final state. In Eq. (\ref{eq:py}) 
$p_{\gamma}$ and $p_{\rho}$ are the 4-momenta of the incoming photon and 
the detected $\pi^+\pi^-$ pair and $p_N$ is the 4-momentum 
of the struck nucleon which, for the calculation of $p_Y$, is assumed to be 
at rest. In addition we introduce a lower boundary for the four-momentum 
transfer $|t-t_{max}|>$0.09~GeV$^2$ as imposed by the HERMES collaboration 
to get rid of coherently produced $\rho^0$. 

\begin{figure}[bt]
  \begin{center}
    \includegraphics[width=6cm]{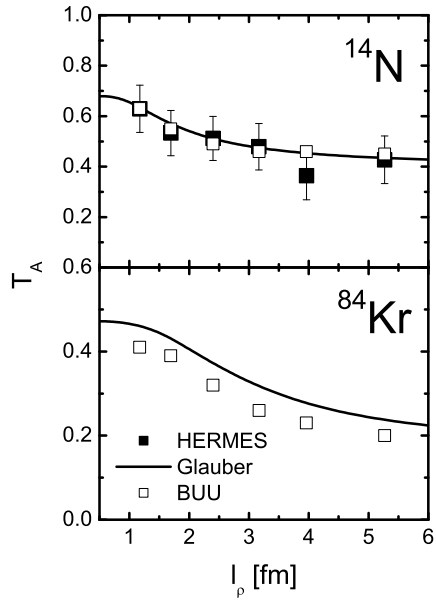}
  \end{center}
  \vspace{-0.5cm}
  \caption{Nuclear transparency ratio $T_A$ for $\rho^0$ electroproduction
plotted versus the coherence length of the $\rho^0$ component of the photon.
The data is taken from~\protect\cite{HERMESrho}.
The solid line represents the Glauber result when using Eq. (\ref{eq:falter}). 
For each transparency ratio calculated within our 
transport model (open squares) we used the average value of $Q^2$ and $\nu$ 
of the corresponding data point. The experimental exclusivity measure (\ref{eq:exclusivity}) has been taken into account for both nuclei.}
\label{fig:fig2}
\end{figure}

In Fig.~\ref{fig:fig2} we show the transparency ratio $T_A$ for exclusive 
$\rho^0$ production as a function of the coherence length 
$l_\rho=q_\rho^{-1}$. The solid line displays the result that one gets if one 
uses our Glauber expression from Ref. \cite{Fal03a}
\begin{eqnarray}
  \label{eq:falter}
  \sigma_{\gamma A\rightarrow \rho^0A^*}&=&\sigma_{\gamma N\rightarrow \rho^0 N}\intop d^2b\intop_{-\infty}^{\infty}dz n(\vec b,z)\nonumber\\
  & &\times\bigg|1-\intop_{-\infty}^{z}dz_i n(\vec b,z_i)(1-j_0(q_c|z_i-z|))\frac{\sigma_{\rho N}}{2}(1-i\alpha_\rho)e^{iq_{\rho}(z_i-z)} \nonumber\\
 & &\times\exp\left[-\frac{1}{2}\sigma_{\rho N}(1-i\alpha_{\rho})\intop_{z_i}^{z}dz_kn(\vec b,z_k)\right]\bigg|^2 \nonumber\\
  & &\quad\times \exp\left[-\sigma_\rho^{inel}\intop_{z}^\infty dz' n(\vec b,z')\right].
\end{eqnarray}
In Eq. (\ref{eq:falter}) we use for the total $\rho^0N$ cross 
section $\sigma_{\rho N}=25$~mb and for the elastic part $\sigma_\rho^{el}=3$ 
mb. These two values correspond to the $\rho N$ cross sections used within the
transport model for the involved $\rho^0$ momenta. 

The result of the transport model is represented by the open squares. For 
each data point we have made a separate calculation with the 
corresponding values of $\nu$ and $Q^2$. In the case of $^{14}$N the Glauber and the 
transport calculation are in perfect agreement with each other and the 
experimental data. This demonstrates that, as we have discussed in 
Ref.~\cite{Fal02}, Glauber theory can be used for the FSI if the right 
kinematic constraints are applied. 

After applying all of the above cuts, nearly all of the detected $\rho^0$ 
stem from diffractive $\rho^0$ production for which the formation time is 
zero. The $^{14}$N data seems to support the assumption that the time needed 
to put the preformed $\rho^0$ fluctuation on its mass shell and let the wave 
function evolve to that of a physical $\rho^0$ is small for the considered 
values of $Q^2$. Furthermore, the photon energy is too low to yield a large 
enough $\gamma$ factor to make the formation length exceed the internucleon 
distance and make color transparency visible. This conclusion is at variance
with that reached in Ref.~\cite{Kop01}.

We now turn to $^{84}$Kr where we expect a stronger effect of the FSI. 
Unfortunately there is yet no data available to compare with. As can be seen
from Fig.~\ref{fig:fig2} the transport calculation for $^{84}$Kr gives a 
slightly smaller transparency ratio than the Glauber calculation, especially 
at low values of the coherence length, i.e. small momenta of the produced 
$\rho^0$. There are two reasons for this: About 10\%~of the difference arises 
from the fact that within the transport model the $\rho^0$ is allowed to decay
into two pions. The probability that at least one of the pions interacts on 
its way out of the nucleus is about twice as large as that of the $\rho^0$. 
The other reason is that in the Glauber calculation (\ref{eq:falter}) only 
the inelastic part of the $\rho^0N$ cross section enters whereas the transport
calculation contains the elastic part as well. Thus all elastic scattering
events out of the experimentally imposed $t$-window are neglected in the
Glauber description. It is because of this $t$-window that also elastic 
$\rho^0N$ scattering reduces the transport transparency ratio shown in 
Fig.~\ref{fig:fig2}. Both effects are more enhanced at lower energies and 
become negligible for the much smaller $^{14}$N nucleus.


\subsection{Hadron formation and attenuation in DIS}
\label{subsec:formation}
As mentioned above hadron production in deep inelastic lepton-nucleus
scattering (DIS) offers a way to study the physics of
hadronization~\cite{Kop1}. The reaction of the exchanged virtual photon
(energy $\nu$, virtuality $Q^2$) with a bound nucleon leads to the
production of several hadrons. Depending on the space-time
evolution of the production process the rescattering in the
surrounding nuclear medium will change the energy distribution and
the multiplicity of the produced particles. Consequently, the
particle spectrum of a lepton-nucleus interaction will
differ from that of a reaction on a free nucleon. In order to
explore such attenuation effects the HERMES collaboration has
investigated the energy $\nu$ and fractional energy $z_h=E_h/\nu$
dependence of the charged hadron multiplicity ratio
\begin{equation}
\label{eq:multiplicity-ratio}
R_M^h(z,\nu)=\left(\frac{N_h(z,\nu)}{N_e(\nu)}\right)_A\bigg/\left(\frac{N_h(z,\nu)}{N_e(\nu)}\right)_D
\end{equation}
in DIS off N~\cite{HERMESDIS_N} and Kr~\cite{HERMESDIS_Kr} nuclei. Here $N_h(z,\nu)$
represents the number of semi-inclusive hadrons in a given ($z,\nu$)-bin and
$N_e(\nu)$ the number of inclusive DIS leptons in the same $\nu$-bin. 

It was suggested in Ref. \cite{HERMESDIS_N}, that a phenomenological
description of the $R_M^h$ data can be achieved if the formation
time, i.e. the time that elapses from the moment when the photon
strikes the nucleon until the reaction products have evolved to
physical hadrons, is assumed to be proportional to $(1-z_h)\nu$ in
the target rest frame. This $(1-z_h)\nu$ dependence of the formation time 
$\tau_f$ is compatible with the gluon-bremsstrahlung model of Ref. 
\cite{Kop}.
In the investigations of Ref. \cite{HERMESDIS_N} any interaction of the reaction 
products with the remaining nucleus during this formation time has been 
neglected. After the formation time the hadrons could get absorbed according to their full hadronic cross section. Another interpretation
of the observed $R_M^h$ spectra -- as being due to a combined effect of a 
rescaling of the quark fragmentation function in nuclei due to partial 
deconfinement as well as the absorption of the produced hadrons -- has
recently been given by the authors of Ref. \cite{Acc02}. Furthermore,
calculations based on a pQCD parton model \cite{Wang,Arl03} explain the 
attenuation observed in the multiplicity ratio solely by partonic multiple 
scattering and induced gluon radiation. It has already been pointed out by 
the authors of Ref. \cite{Acc02} that a shortcoming of the existing models 
is the purely absorptive treatment of the final state interactions (FSI). We avoid this problem by using the coupled-channel transport
model.

In our calculation we employ the kinematic cuts of the HERMES
experiment as well as the geometrical cuts of the detector. In
actual numbers: we require for the Bjorken scaling variable
$x=\frac{Q^2}{2m_N\nu}>0.06$ (with $m_N$ denoting the nucleon
mass), for the photon virtuality $Q^2>1$~GeV$^2$ and for the
energy fraction of the virtual photon $y=\nu/E_{beam}<0.85$. In
addition, the PYTHIA model introduces a lower cut in the invariant
mass of the photon-nucleon system at $W=4$~GeV that is above the
experimental constraint $W>2$~GeV. This limits our calculations to
minimal photon energies of $\nu_{min}=8.6$~GeV as compared to
$\nu_{min}=7$~GeV in the HERMES experiment and leads to a
suppression of high $Q^2$ events at energies below
$\nu\approx 15$~GeV. 

We have shown in Ref. \cite{Fal03b} that our model simulation reproduces the experimental average values of the kinematic variables $Q^2$ and $\nu$ ($z_h$) on {\it N} and {\it Kr} as a function of $z_h$ ($\nu$) very well. Since the particles with $z_h$ close to 1 are predominantly leading hadrons we could use the high $z_h$ part in the fractional energy spectrum
to fix the leading hadron cross section. A good agreement with the data is 
achieved for $\sigma_{lead}=0.33\sigma_h$ during the formation time
$\tau_f$. We note, that this value for
$\sigma_{lead}$ represents an average value over time from the virtual 
photon-nucleon interaction to the actual hadron formation time. 
Since for light nuclei only a fraction of the leading hadrons is 
formed inside the nucleus, the effective leading hadron cross section 
$\sigma_{lead}$ is expected to be smaller accordingly. For a detailed 
investigation we refer the reader to a forthcoming study~\cite{F2}.

\begin{figure}[bt]
  \begin{center}
    \includegraphics[width=8.5cm]{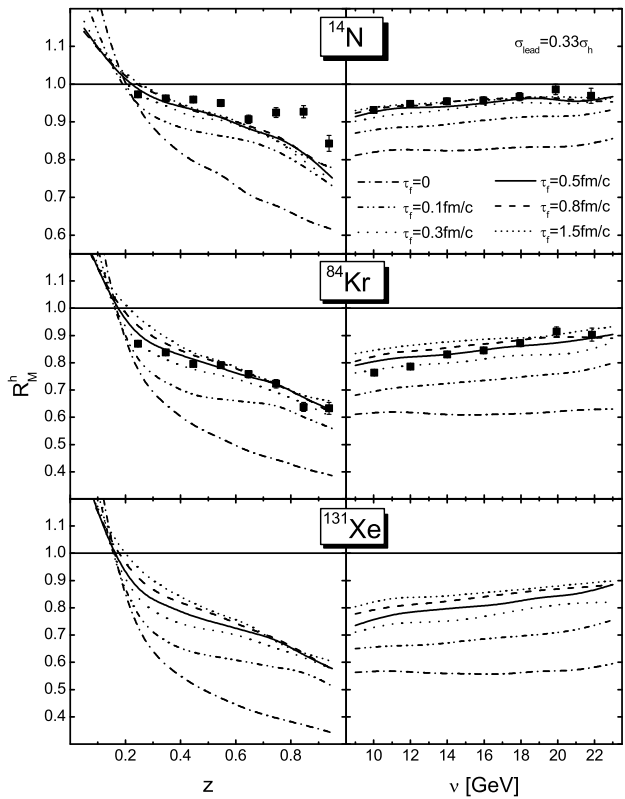}
  \end{center}
  \vspace{-0.5cm}
  \caption{Calculated multiplicity ratios of charged hadrons for {\it N}, {\it Kr} and {\it Xe} 
targets for a fixed leading hadron cross section 
$\sigma_{lead}=0.33\sigma_h$ and different values of the formation time 
$\tau_f$.
The data for the Nitrogen target have been 
taken from Ref. \protect\cite{HERMESDIS_N} while the preliminary {\it Kr} data stem from 
Ref. \protect\cite{HERMESDIS_Kr}.}
\label{fig:fig3}
\end{figure}

In Fig.~\ref{fig:fig3} we show the influence of different formation
times $\tau_f$ on $R_M^h$ using the effective cross section 
$\sigma_{lead}=0.33\sigma_h$. We find, that formation times 
$\tau_f\gtrsim 0.3$ fm/c are needed to describe the experimental data with 
little sensitivity to higher values. This is compatible to the range of 
values extracted from the antiproton attenuation studies in Ref. 
\cite{Cas02}.

Some of our model assumptions, e.g. the local density approximation, become
questionable for very light nuclei. We have, therefore, used the {\it Kr} data 
to fix the value of the formation time. However, we also get a satisfying 
agreement with the Nitrogen data.

The pQCD model of Ref. \cite{Wang} predicts a hadron attenuation 
$\sim A^{2/3}$ since the parton energy loss is
proportional to the propagation length squared. It is thus
important to get the scaling with target mass $A$ from our present
approach in order to allow experimental studies to distinguish
between the different concepts. To this aim Fig~\ref{fig:fig3}
also shows predictions for a Xe target. In accordance with the
authors of Ref.~\cite{Acc02} we predict only a small change in the
multiplicity spectra compared to the {\it Kr} target such that the 
scaling exponent is lower than $2/3$.

\begin{figure}[bt]
  \begin{center}
    \includegraphics[width=7.9cm]{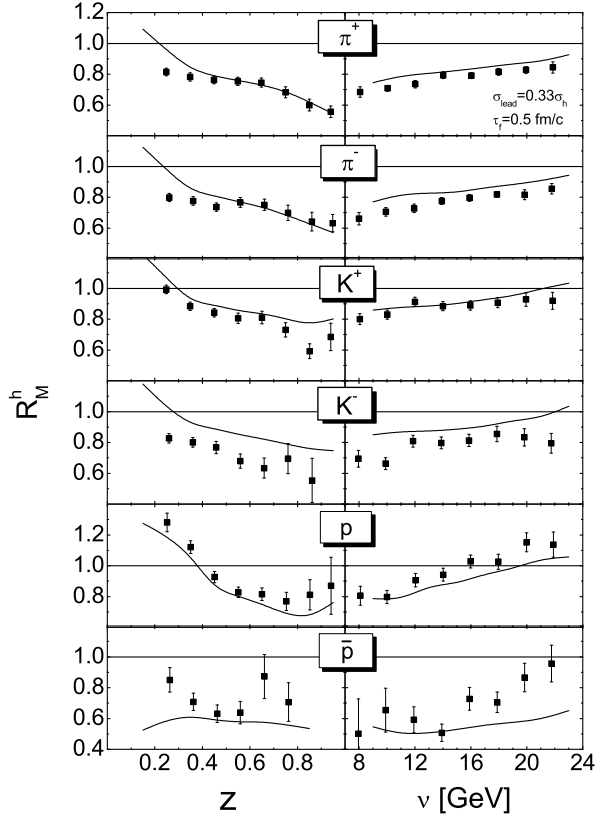}
  \end{center}
  \vspace{-0.5cm}
  \caption{Calculated multiplicity ratios of $\pi^+$, $\pi^-$, $K^+$, $K^-$,
$p$ and $\bar{p}$ for Krypton using a fixed leading hadron cross section 
$\sigma_{lead}=0.33\sigma_h$ and formation time $\tau_f=0.5$~fm/c.
The experimental data has been taken from Ref. \cite{HERMES_new}.}
\label{fig:fig4}
\end{figure}

In Fig. \ref{fig:fig4} we show the results for the calculated multiplicity ratio of $\pi^-$, $\pi^+$, $K^-$, $K^+$, $p$ and $\bar{p}$ for {\it Kr} in comparison with the experimental data. In our calculations we use again the kinematic cuts of the HERMES experiment with the additional restrictions $x\geq 0.02$, $E_{\pi,K}$=2.5--15 GeV and $E_{p,\bar{p}}$=4--15 GeV that are experimentally necessary for particle identification \cite{HERMES_new}. We use a constant formation time of 0.5 fm/c and $\sigma_{lead}=0.33\sigma_h$ for all hadrons. Without further fine tuning we get a satisfying description of all the data meaning that the formation times of mesons, baryons and antibaryons are about equal.


\section{Summary}\label{sec:summary}
We have developed a method to account for coherence length effects within a semi-classical BUU transport model. This allows us to describe  
incoherent meson photo- and electroproduction off nuclei at GeV energies using a full coupled channel treatment of the FSI. We have calculated the transparency ratio for exclusive incoherent 
$\rho^0$ photoproduction off $^{14}$N and $^{84}$Kr. The result for $^{14}$N is
in agreement with experimental data and with the Glauber prediction. The 
latter shows that in the case of $^{14}$N Glauber theory is applicable after 
the kinematic cuts of the HERMES experiment are applied. Since we do not use a 
formation time for diffractively produced vector mesons we deduce that one 
cannot see an onset of color transparency in the Nitrogen data. For the 
$^{84}$Kr target no experimental data is available to compare with. 
However, we find deviations from the simple Glauber model because of the 
finite life time of the $\rho^0$ and elastic scattering out of the 
kinematically allowed $|t|$-region.
As discussed by Kopeliovich et al.~\cite{Kop01} one might see an onset of color transparency when investigating the transparency ratio as a function of $Q^2$ for fixed coherence length. 
 
In addition we have shown that one can describe the experimental data of the 
HERMES collaboration for hadron attenuation on nuclei without invoking any 
changes in the fragmentation function due to gluon radiation. 
In our dynamical studies, that include the most relevant FSI, we employ only 
the 'free' fragmentation function on a nucleon and attribute the hadron 
attenuation to the deceleration of the produced hadrons due to FSI in the 
surrounding medium. We find that in particular the $z$-dependence of $R_M^h$ 
is very sensitive to the interaction cross section of leading hadrons and 
can be used to determine $\sigma_{lead}$. The interaction of the leading 
hadrons during the formation time could be interpreted as an in medium change 
of the fragmentation function. The extracted average hadron formation times 
of $\tau_f \gtrsim 0.3$ fm/c are compatible with the analysis of antiproton 
attenuation in $p+A$ reactions at AGS energies \cite{Cas02}. 

In this work we have also for the first time compared our results for the attenuation of $\pi^-$, $\pi^+$, $K^-$, $K^+$, $p$ and $\bar{p}$ with the experimental {\it Kr} data. We find a satisfying description of all data using the same constant formation time $\tau_f$=0.5 fm/c and the same reduction of the effective leading hadron cross section $\sigma_{lead}$=0.33 fm/c for all hadron species.


\section*{Acknowledgments}
This talk is based on work done together with W. Cassing and K. Gallmeister.
The authors acknowledge valuable discussions with A. Borissov, C. Greiner and
V. Muccifora. This work was supported by DFG and BMBF. 


\end{document}